# The Multiple Systems in The Young Stellar Cluster IRAS 05137+3919

**E. H. Nikoghosyan, H. A. Harutyunyan, N. M. Azatyan**

*Byurakan Astrophysical Observatory, Armenia, elena@bao.sci.am*

***Abstract.*** Four binary objects and one triplet have been revealed in the young stellar cluster located in the vicinity of IRAS 05137+3919 source on a distance 4.4 kpc with the use of statistic analysis. They are including the pair of Ae/Be Herbig stars. The percentage of the multiple systems in the cluster is mf = 5-6% and cp = 10-13%. The mass of the multiple systems' components are located in the range from ~ 1 to 8 $M_{\odot}$ and log P (rotation period in years) - from 4.4 to 4.7. The median value of the mass ratio of the components is q=0.86.

The percentage of the multiple systems and their parameters in this cluster is resembling with the data obtained in the other star forming regions (ONC, Perseus, U Sco A), in which the values of mf and cp parameters are comparable with the results obtained for field's stellar population.

***1. Introduction.*** Nowadays, the theory of star formation has generally recognized two aspects: 1) the stellar population is born in associations (Ambartsumyan, 1947; Lada & Lada, 2003); 2) the most stars are born in binary or multiple systems (Ambartsumyan, 1947; Duchene & Kraus, 2013). Therefore, the study of multiple systems, both in young star clusters and among the objects of the field is one of the main directions of modern astrophysical research. From the viewpoint of the foregoing, it is logical to expect that there should be a definite correlation between the percentage of multiple systems in the field and young clusters. However, many studies show that in the number of clusters, such as the Taurus, Ophiuchus, Chamaeleon, Lupus, Corona Australis, percentage of multiple systems is considerably higher than in the field (Duchene & Kraus, 2013 and ref. therein). At the same time, in such regions as ONC, Perseus, U Sco percentage of binary stars, on the contrary, is comparable to the field, and even less (Duchene & Kraus, 2013 and ref. therein). There are different explanations regarding to such differences. The first explanation is following: some parts of multiple systems in young clusters are being destructed before its components reach the Main Sequence (Ghez et al, 1993). There is also an alternative assumption - the percentage of multiplicity of PMS and MS stars should be the same, but should evolve rotation periods (Reipurth & Zinnecker, 1993) and the majority of the field stars are born in rich young clusters with high density, such as ONC, rather than in relatively poor T-associations, such as Tau-Aur (Reipurth & Zinnecker, 1993; Bouvier et al, 1997). In clusters themselves, percentage and distribution of rotation periods depend on the certain initial conditions (Connelley et al, 2008) such as size of the star forming region (Kouwenhoven et al, 2010) or a temperature of ambient gas and dust (Kraus & Hillenbrand, 2007). It should also be noted, that because of the significant differences between the parameters (resolution, depth) of the observation data, there are certain difficulties in comparing the already obtained results. In all cases, these issues require further detailed study.

The main subject of our research are 20 compact star forming regions located in the vicinity of high and intermediate mass YSOs (Varricatt et al, 2010 and ref. therein). In these areas, there are different manifestations of activity, including molecular outflow (CO and $H_2$), $NH_3$, $H_2O$ and $CH_3OH$ emission and etc. All these areas are covered by GPS UKIDSS survey. The purpose of research is not only to determine and define the parameters of multiple systems, but also the study of the results of static analysis, depending on the used databases, namely: 2MASS and GPS UKIDSS.

To identify the multiple systems we use Poisson statistics and a two-point correlation functions (2PCF, TPACF). Since, the foregoing regions are located at relatively large distances and we can not use the proper motions of objects.

In this work we present the results of the multiple systems' study of R-associations located in the vicinity of IRAS 05137 + 3919 source. The radius of the cluster is 1.5 arcmin and distance is 4.4 kpc. (Nikoghosyan & Azatyan, 2014).

## *2. The research method.*

***2.1. The observational data.*** To identify and study of binary stars in the cluster were used images, coordinates and photometric data taken from 2MASS (Skrutskie et al, 2006) and GPS UKIDSS (Lucas et al, 2008) IR surveys. To identify binary stellar objects were used K band images. The astrometric accuracy of both surveys is 0.1 arcsec. The image resolution is 1″ and 0.42″ per pix respectively. For the purity of the sample with respect to the photometric parameters we have selected objects with K>15.4 and K > 18.05 of the 2MASS and GPS UKIDSS databases respectively. At a distance of 4.4 kpc the sample from the 2MASS survey includes only stars with spectral classes no later than F0, from GPS UKIDSS - no later than M0. In addition, in the second case the objects for which the probability that they are the result of different kinds of noise or defects of images is more than 90% were excluded. From both lists were also excluded knots of outflows revealed in the $H_2$ spectral lines (Varricatt et al, 2010).

To reveal close binaries those in the foregoing databases were identifying as single extended source, we have analyzed all objects from 2MASS survey with «Extended source contamination» index 2 and all objects from the GPS UKIDSS survey with the ellipticity of more than 0.2. By the close binary objects were assigned the extended sources, in which two well defined maximums of brightness are noticeable. An example of such an object, which surface brightness on the K image from the GPS UKIDSS has two well-defined maximum, but identify as a single object with the ellipticity 0.2, is shown in Fig. 1. Thus, 10 and 6 close binaries were selected from the 2MASS and the GPS UKIDSS lists respectively. Their astrometric and photometric parameters were determined by using the NOAO IRAF/DAOPHOT software package. It should noted, that the accuracy of photometry in this case is much lower and is ~ 0.2 m. This analysis also allowed to exclude the nebulous objects from the final list, namely, clumps of gas and dust matter and galaxies.

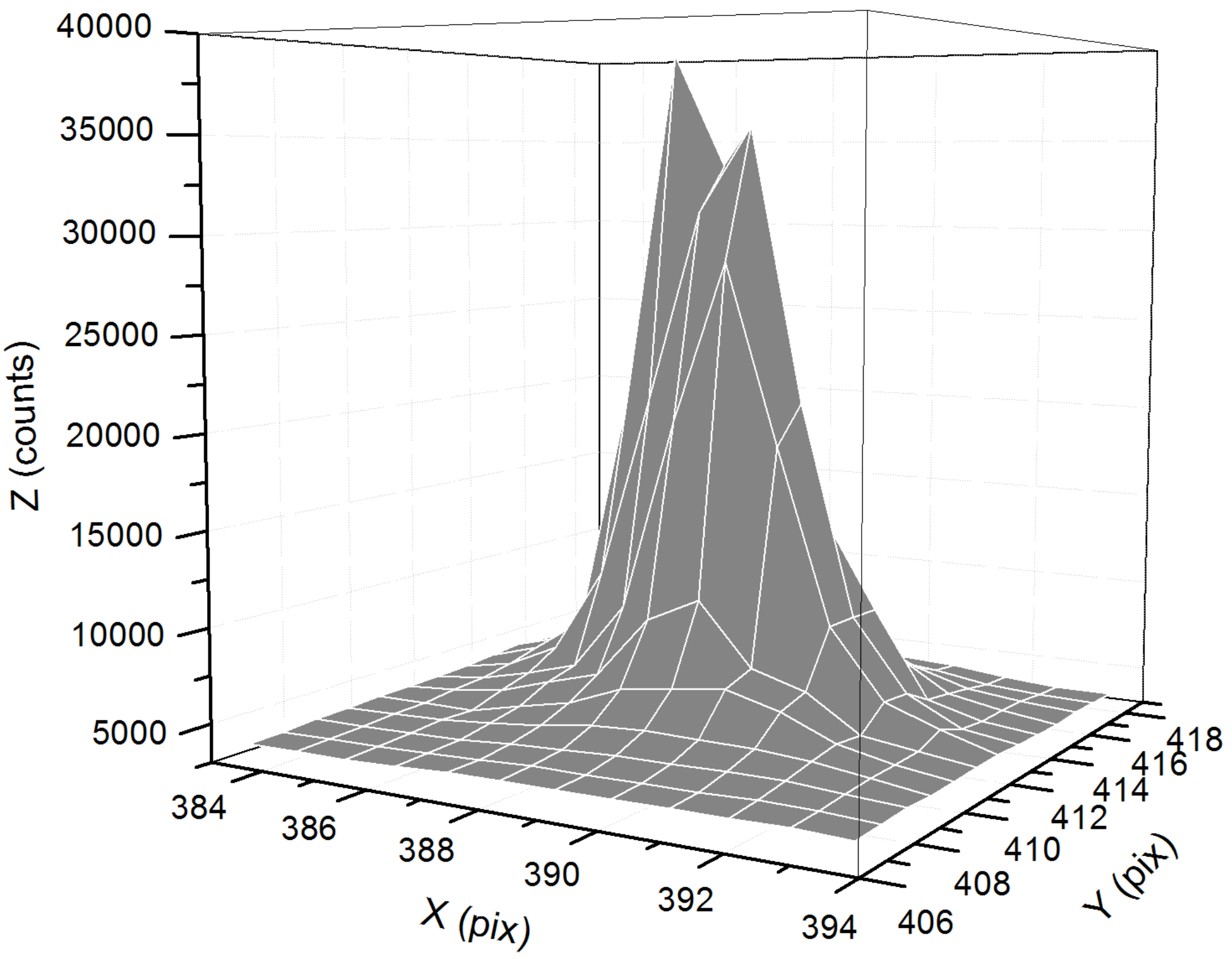

**Fig. 1.** *The distribution of surface brightness of the close binary object.*

A total from 2MASS database were identifying 49 stellar objects, from the GPS UKIDSS - 272.

Furthermore, the images in the K and $H_2$ bands taken from the work (Varricatt et al, 2010) were used.

***2.2. Statistical analysis.*** For the identification of binaries and triplets in the cluster was used statistical analysis of the distribution of interstellar distances and determined the degree of its deviation from random. For this purpose, we used three statistical test.

*Two-point correlation function* (2PCF). 2PCF function or ω(θ) determines the observed excess of pairs, spaced at an angular distance θ relative to a random distribution:

$$\omega(\theta) = \frac{F(\theta)}{P(\theta)} - 1,$$ (Peebles, 1980; Longhitano & Binggeli, 2010),

where F(θ) the number of observed stellar pairs with an angular separation between θ and θ + dθ and $P(\theta)d\theta \approx \frac{\pi N_{obs}(N_{obs}-1)}{\Omega}\theta d\theta$ – the number of pairs expected from a random distribution in the sample with $N_{obs}$ points, Ω - the solid angle of the region.

*Two Point Angular Correlation Function (TPACF).* This test compares the distribution of angular separations between objects in the real cluster at the N-th number of random generated samples with the same number of objects and the coordinates intervals (Kerscher et al, 2000):

$$\omega(\theta) = \frac{DD(\theta) - 2\frac{\sum DR(\theta)}{N} + \frac{\sum RR(\theta)}{N}}{\frac{\sum RR(\theta)}{N}},$$

where DD and RR - are the auto-correlation functions of the actual and generated data, DR - the cross-correlation function of the actual data set to a given random set, N - a number of the random sets.

*Poisson statistics.* According to Poisson statistics the probability of finding k stars within distance r of another star is

$$W(k,n) = \frac{n^k}{k!}e^{-n}$$ (Petr et al, 1998),

where n (a number of objects in the sample) = $\pi r^2 n_*$, and $n_*$ - surface density. Thus (taking into account, that $e^{-a} \approx 1 - a \approx 1$, then $a \ll 1$), the probability that from a given object on the distance r to be located only one or two objects will $W = \pi r^2 n_*$ and $W = 0.5\pi^2 r^4 n_*^2$ respectively.

For the sample with N objects expected number of random pairs ($N_p$) and triplets ($N_T$) will $N_p = N\pi r^2 n_*$ and $N_T = 0.5N\pi^2 r^4 n_*^2$ respectively.

It should be noted that, since we are study a young star cluster, an important factor in the selection of binary stars are also their photometric parameters, based on which we can do certain conclusions about their evolutionary stage and, therefore, belonging to the cluster.

## *3. Results.*

***3.1. Statistical analysis of data.*** Described in the previous section, statistical methods have been consistently applied to different samples of stellar objects. These samples include at first the list the 84 PMS members of the cluster (Nikoghosyan & Azatyan, 2014). The revealing of close pairs increases this number up to 88, representing 76% of the excess of stellar objects relative to the ones in the field in the vicinity of the cluster (Nikoghosyan & Azatyan, 2014). Besides, the objects taken from both the 2MASS and the GPS UKIDSS with magnitudes up to K ≤ 15.5. Note that at a distance of 4.4 kpc it is practically corresponds to the main sequence stars with spectral classes B and A or PMS objects with mass greater than ~1.4 $M_{\odot}$ (Patience & Ghez, 2002). Because according to the data given in Nikoghosyan & Azatyan, 2014 $A_k$ does not exceed 0.2 m the interstellar absorption is not taken into account. The next sample - the objects taken from the GPS UKIDSS with magnitudes from 15.5 to 18.5, which corresponds to ZAMS spectral classes from F0 to M0 or PMS stars with masses from 1.4 $M_{\odot}$ to 0.6 $M_{\odot}$. In addition, the statistical analysis was also applied to all objects from the GPS UKIDSS.

Fig. 2 and 3 show the plots of the dependence of ω (θ) from the angular distance between objects, (obtained by using the 2PCF and TPACF respectively). According to the dependence of ω(θ) on θ on both Fig. 2 and Fig.3, it is clearly shows that, among the brightest objects and PMS stars there is an excess of pairs (relative to a random distribution), located at a distance of less than 0.6″. This is considerably less than the value r (r = 1.0″ (PMS stars), r = 1.8″ (2MASS) and r = 1.5″ (UKIDSS K≤15.5), see sec. 2.2), at which, according to Poisson statistics, the number of randomly projected pairs is minimally, ie equal to 1. Even if we take into account the fact that, as mentioned in (Nikoghosyan & Azatyan, 2014), the excess of stellar objects in cluster relative to field is 115, i.e., identified PMS stars is only about 76% of total amount, and in this case r will be equal to 0.78".

For cluster as a whole, as well as for the faint stars (K < 15.5), such excess is not observed.

Consistent with the above, we have selected binary stars, which angular separations are less than 0.6″. Their coordinates, magnitudes and the projection of the distance between them are shown in Table 1. The distance between each member of the pairs to the nearest second object exceeds their separation more than three times. Except four binaries, it was selected the triplet in which the mutual distance between stars is less than a maximum value given by Poisson statistics (for sample of 88 objects r = 3.7″, for 115 - r = 3.0″). Note, that the distance between each member of the triplets to the nearest third object also exceeds the maximal separation more than three times. The presence of a triplet also reflects the plot on Fig. 2 and 3, where in the interval of θ from 2 to 2.5 arcsec there is some excess of objects. It should be noted that the 2MASS data revealed only 4 - th pair from Table 1 and two objects in the triplet.

Images of objects from Table 1 are shown in Fig. 4.

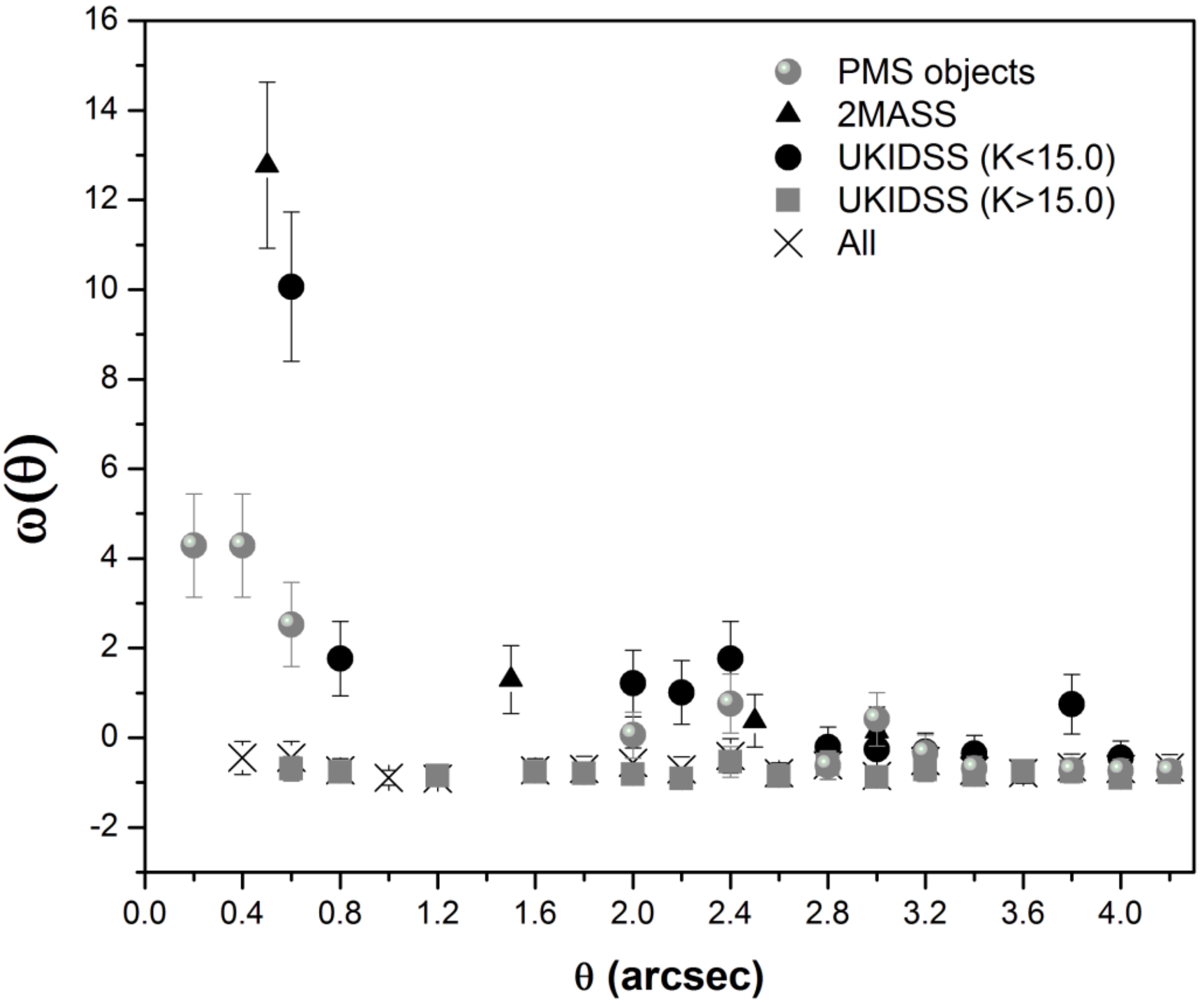

***Fig. 2***. *The dependence of ω (θ) from angular separation θ, according to the 2PCF.*

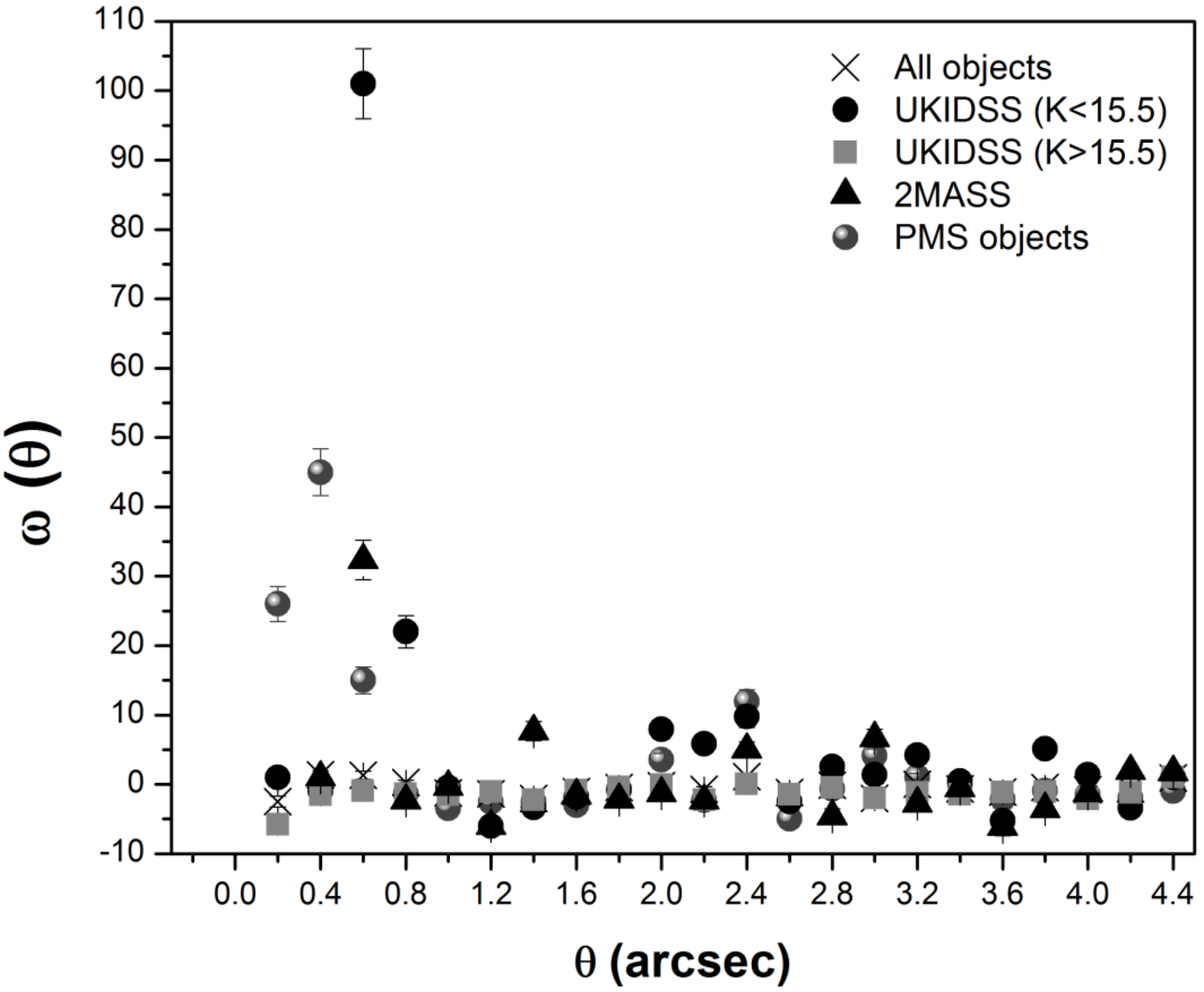

***Fig. 3***. *The dependence of ω (θ) from angular separation θ, according to the TPACF.*

*Table 1*

## Binary Systems and Triplet

| N | RA (2000) | Dec (2000) | d (arcsec, AU) | J | H | K |
|---|---|---|---|---|---|---|
| **Binary Systems** | | | | | | |
| 1 | a) $5^d\ 17^m\ 08.41^s$ | 39° 21′ 29.2″ | 0.31 | 16.72 | 15.53 | 14.50 |
| | b) 5 17 08.42 | 39 21 29.4 | 1364 | - | 16.49 | 15.57 |
| 2* | a) 5 17 11.13 | 39 21 46.2 | 0.31 | - | 16.37 | 16.02 |
| | b) 5 17 11.15 | 39 21 46.0 | 1364 | 17.42 | 15.97 | 15.09 |
| 3 | a) 5 17 12,69 | 39 21 55.4 | 0.44 | 18.06 | 16.26 | 15.48 |
| | b) 5 17 12.71 | 39 21 55.8 | 1936 | 17.34 | 16.31 | 15.49 |
| 4* | a) 5 17 13.70 | 39 22 19.0 | 0.50 | 13.50 | 12.2 | 10.8 |
| | b) 5 17 13.68 | 39 22 19.4 | 2200 | - | - | 10.6 |
| **Triplet** | | | | | | |
| 1 | a) 5 17 13.44 | 39 21 52.6 | a-b 1.85, 8156 | 14.55 | 14.00 | 13.48 |
| | b) 5 17 13.36 | 39 21 54.2 | b-c 2,23, 9821 | 15.33 | 14.75 | 14.25 |
| | c) 5 17 13.55 | 39 21 54.4 | a-c 2.21 9736 | 15.98 | 14.96 | 14.52 |

(*) – by the asterisks marked the stars which in databases identifying as a single object.

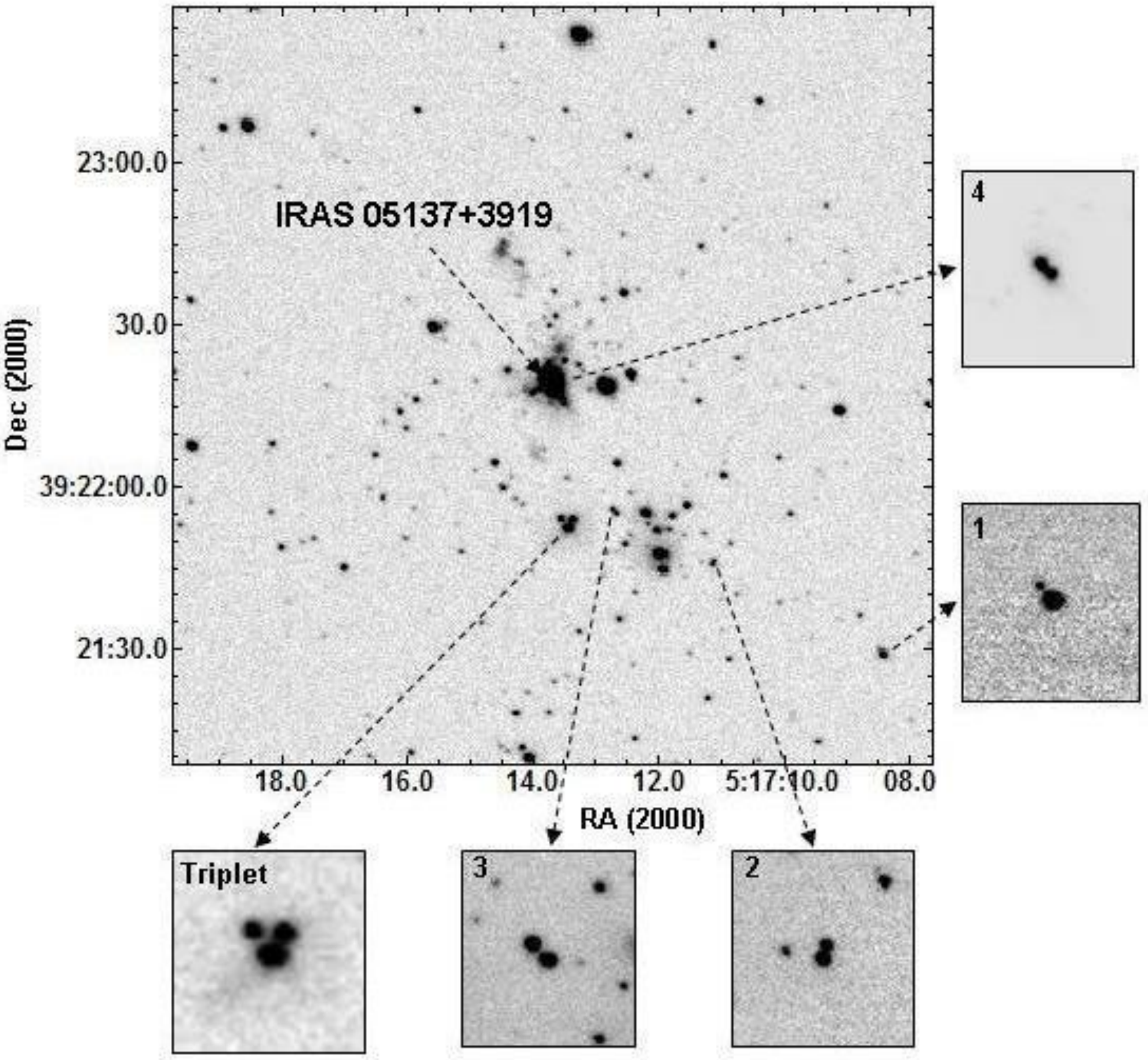

**Fig. 4.** *The images of the binary stars and triplet (K band).*

***3.2. The parameters of the multiple systems.*** Fig. 4 shows the JH/HK two-color diagram on which the positions of the stars from Table 1 are marked. By their position on the diagram we can conclude that the vast majority of stars, for which have been identified JHK magnitudes, can be classified as PMS stars. Therefore, they are likely to belong to the cluster and their relative position is not the result of projection, i.e. these objects can be considered as real pairs and triplet.

Using the relation between the absolute magnitude and mass for yong stars (Patience & Ghez, 2002) we have determined the mass of these objects with the cluster distance 4.4 kpc and absorption Av = 1.8 m (Nikoghosyan & Azatyna, 2014). The results are presented in Table 2. In Table 2 also presented the periods of binary systems, defined according to the relation $a^3/P^2 = m_{sys}$ , where $a$ is a semi-major axis of the orbit in AU, P is a rotation period in years and $m_{sys}$ is a mass of binary system (Goodwin et al, 2006). Of course, it should be noted that in the calculation, we have used the projection of the distance between objects, rather than the actual value of semi-major axis. However, calculations show that with a uniform distribution of the planes of the orbits of binary stars in the sky, between the projection distance ρ and semi-major axis of the orbit $a$ there is the following relationship $\bar{\rho} = 0.85a$ (Couteau, 1960; Reipurth et al, 2007). This, especially taking into account the errors in determining the coordinates of close binaries, makes only a very minor adjustment to the final result.

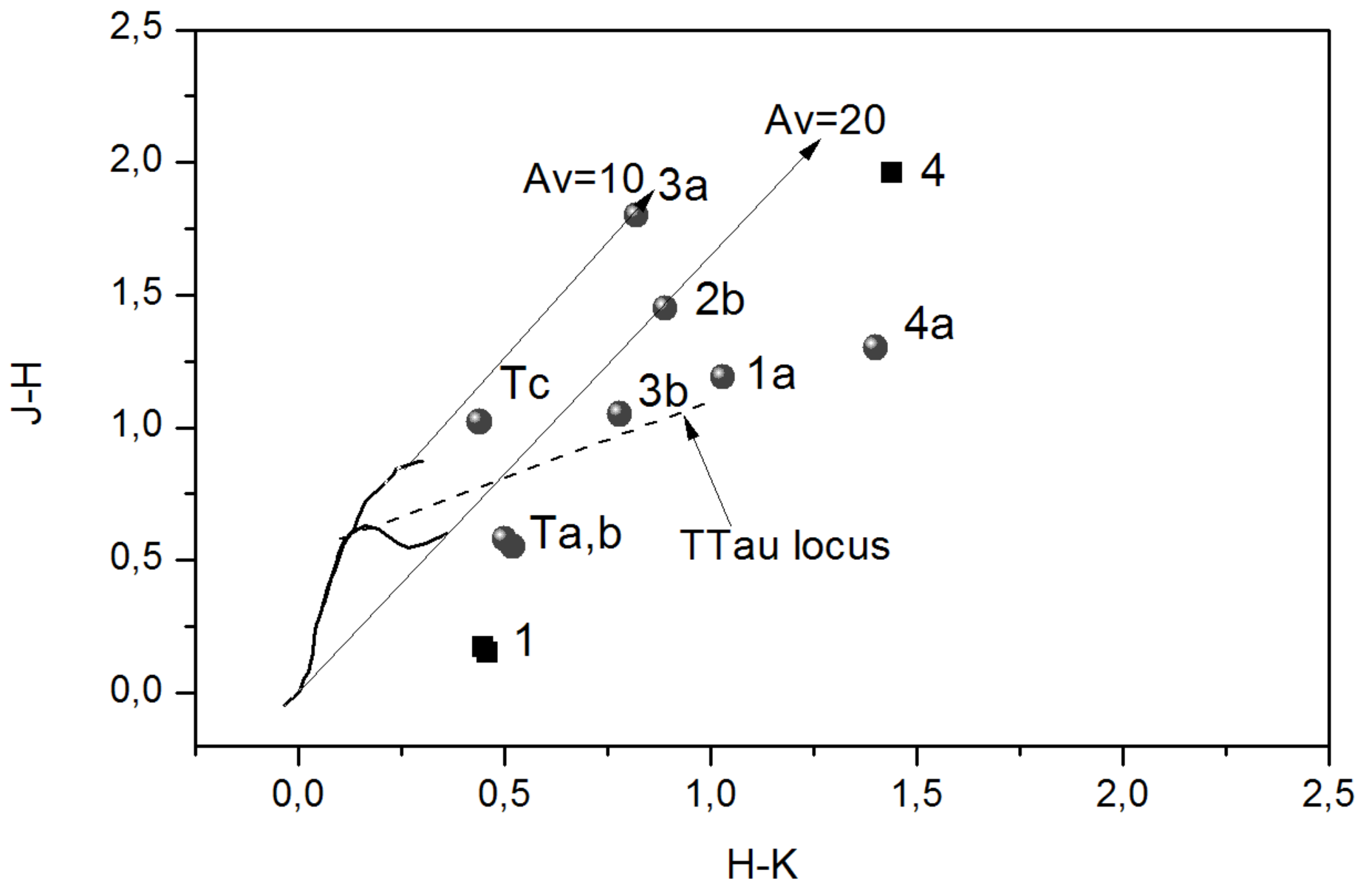

**Fig. 4.** *JH/HK two-color diagram. The circles mark the positions of the stars from the Table 1 preserving the notation (a number of binary system, T - triplet); squares - photometric parameters of originally unresolved object from GPS UKIDSS database. The MS and GB are taken from Bessell & Brett (1988), TTau locus - from Meyer & Calvet (1997), reddening vectors – from Cardelli et al (1989).*

*Table 3.*

**The Masses and Periods of the Multiple Systems**

| Objects | Mass (⊙) | Log P (year) |
|---|---|---|
| **Binary Systems** | | |
| 1 a | 1.95 | 4.44 |
| 1 b | 1.32 | |
| 2 a | 1.12 | 4.48 |
| 2 b | 1.57 | |
| 3 a | 1.36 | 4.71 |
| 3 b | 1.36 | |
| 4 a | 7.55 | 4.42 |
| 4 b | 8.12 | |
| **Triplet** | | |
| a | 2.83 | - |
| b | 2.13 | |
| c | 1.93 | |

Consider each of the systems in the Table 1 in detail.

***Pair 1***. Initially, this binary object was defined as a pair of MS stars, practically from the same luminosity. Their photometric parameters are not provided a basis to classify these as young stars, so they were not included in the list of PMS objects in (Nikoghosyan & Azatyan, 2014). However, K band image adopted from Varricatt et al, 2010 clearly shows that the components of the pair significantly differ in brightness, so we revised magnitudes of these objects. The most luminous of them on JH/HK diagram is located in the vicinity TTau locus. The second with respect to its color H - K = 0.92 can also be considered as PMS stars.

***Pair 2.*** In (Nikoghosyan & Azatyan, 2014) this object is classified as PMS star. However, more detailed analysis showed that it is binary object, where the second component is inferior to the first by brightness and invisible in J band. Based only on the color index H-K = 0.35 difficult to reliably judge about the evolutionary status of this star: is it PMS object and belongs to cluster or it is the result of the projection. Note that in this case, the probability of random projection is only ~ 0.3%.

***Pair 3.*** This is a pair of already known young stars (Nikoghosyan & Azatyan, 2014), located in the vicinity of the secondary maximum of a continuum emission in the submillimeter (SCUBA) and infrared (MSX) ranges, as well as HCO + and CS emission (Molinary et al, 2002).

***Pair 4.*** This pair of Ae/Be Herbig stars, which is associated with YSO CPM 15 or IRAS 05137 + 3919 source (Nikoghosyan & Azatyan, 2014). Both components of this binary are sources of collimated bipolar outflow identified in the spectral lines $H_2$ v =1–0 S(1) (2.1218 μm) and Br (2.166 μm) (Varricatt et al, 2010).

***Triplet.*** This group of stars is surrounded by a well recognizable in MID spherical nebula (Nikoghosyan & Azatyan, 2014). The half-life in years for the triplets can be defined by the formula $t = 7\left(\frac{R}{au}\right)^{3/2}\left(\frac{M_*}{M_{sol}}\right)^{-1/2}$, where R is the radius of the cluster and $M_*$ is the mass of the components (Anosova, 1986). In this case $t \approx 2 \cdot 10^6$, is comparable with the age of cluster (Nikoghosyan & Azatyan, 2014).

**4. *Discussion.*** Through statistical analysis of the data from 2MASS and GPS UKIDSS in relatively distant (~ 4.4 kpc) young star cluster located in the vicinity of IRAS 05137 + 3919 source, we were able to identify four binary stars and one triplet. Practically all stars in these systems based on their photometric data can be classified as PMS objects, which greatly increases the likelihood that they belong to the cluster. Consequently, they can be regarded as real, physically connected systems. In this way the multiplicity frequency $mf = \frac{B+T}{S+B+T}$ (S – a number of single stars, B – a number of pairs, T – a number of triplets) for PMS objects is 6% (Reipurth & Zinnecker, 1993). However, if we take into account that according to the results presented in (Nikoghosyan & Azatyan, 2014), the excess of the stellar objects in the area of the cluster relative to objects in the field is 115 the percentage decreases to 5%. Similarly, the companion probability $cp = \frac{2B+3T}{S+2B+3T}$ (Reipurth & Zinnecker, 1993) are 13% and 10% for both cases. Notice, that for 2MASS selection mf = 4% and cp = 8%. Our values for the parameters mf, and cp are in good agreement with the results obtained for the stars with masses from 1.1 to 1.3 $M_\odot$ in U Sco A association (Kraus & Hillenbrand, 2007), for objects in Perseus cluster (Patience et al, 2002; Haisch & Greene, 2004), as well as for ONC (Reipurth et al, 2007). But, it should be noted that in (Patience et al, 2002) and (Reipurth et al, 2007), this result was obtained for binaries with separations from 20 AU to 700 AU. Here the pairs are much wider. It can be expected that with increasing image resolution, the number of detected multiple systems will also increase.

Periods of the binary stars coincide with the maximum of the distribution of periods in other young clusters, for example, Trapezium (Kroupa et al, 1999) or Pleiades (Bouvier et al, 1997), as well as with A stars in a field (Parker & Meyer, 2014).

We would like to draw your attention to one fact. Using a database of the GPS UKIDSS, we were able to identify the close binary objects, the brightness with lower luminosity than 2MASS limit. However, the faintest components of binary systems has K ≈16.0, which corresponds to an object with a mass ~ 1.1 $M_\odot$. But, we could not find binary systems in which both components or only one of them would have been the stars of lower mass. In the first case, this fact is difficult to explain by the selectivity. Furthermore, as the results of research of many authors, the values of mf and cp decrease with decreasing mass of stars (Duchene & Kraus, 2013). However, in the second case - it can be assumed that the absence of such pairs is the result of selection, since relatively close pairs easier to detect then both components have comparable masses, than when they are significantly different. As noted in a number of studies, such selection leads to certain difficulties in the construction of the distribution of the mass ratio q of binaries (Duchene & Kraus, 2013 and ref. therein). However, in our case, the value of $q = M_2/M_1$ is in the range of 0.67 to 1 with a median value of 0.86, that is some higher, than the results obtained for the four regions of star formation (Taurus-Auriga, Chamaeleon I, Upper Scorpius A and B), where the maximum of q parameter falls on the interval 0.6 - 0.7 for binaries with masses from 1.2 to 2. 5 $M_\odot$ (Kraus & Hillenbrand, 2007).

***5. Conclusion***. Statistical analysis of stellar populations in young cluster located in the vicinity of IRAS 05137 + 3919 source revealed four binary stars and one triple. All stars, except for one object, in their photometric parameters are more likely to be classified as PMS objects. Therefore, identified multiple systems, with high probability, belong to the cluster and they are not the result of a projection. Among them is a pair of Ae/Be Herbig stars, which, moreover, is the source of collimated outflow. The multiplicity frequency in cluster is mf = 5 - 6 %, and companion probability is cp = 10 - 13 %. The masses of the components in multiple systems are in the range from 1 to 8 $M_\odot$**.** We were unable to detect binary stars with masses less than solar. The median value of $q = M_2/M_1$ is 0.86.

The percentage composition of multiple systems and their parameters are similar to those obtained for other, much nearer located regions of star formation ONC, Perseus, U Sco A, in which the values of mf and cp are similar to the results obtained for the stellar population of the field (Duchene & Kraus, 2013).

This work was supported by a grant from the State Committee of Science of Armenia 13-1C087.

We thanks the creators of 2MASS и UKIDSS Galactic Plane Survey databases, as well as the authors of work Varricatt et al (2010) for vested the possibility to use the images.